\documentclass{aa501}
\usepackage{psfig}

\begin{document}

\title{A 1.2~mm MAMBO/IRAM-30m Survey of Dust Emission from the Highest Redshift PSS Quasars}

\author{Alain Omont\inst{1}
       \and Pierre Cox\inst{2}
       \and Frank Bertoldi\inst{3}
       \and Richard G. McMahon\inst{4}
       \and Chris Carilli\inst{3,5}
       \and Kate G. Isaak\inst{6}
        }

\offprints{A. Omont, omont@iap.fr}

\institute{
     Institut d'Astrophysique de Paris, CNRS, 98bis boulevard Arago, F-75014 Paris, France
\and Institut d'Astrophysique Spatiale,  Universit\'e de Paris XI, F-91405 Orsay, France
\and Max-Planck-Institut f\"ur Radioastronomie, Auf dem H\"ugel 69, D-53121 Bonn, Germany
\and Institute of Astronomy, Madingley Road, Cambridge CB3 0HA, UK
\and National Radio Astronomy Observatory, P.O. Box O, Socorro, NM 87801, USA
\and Cavendish Laboratory, Madingley Road, Cambridge CB3 0HE, UK
           }

\date{Received date / Accepted date}

\titlerunning{A 1.2~mm MAMBO survey of the highest redshift PSS quasars}
\authorrunning{A. Omont et al.}

\abstract{
  We report 250~GHz (1.2~mm) observations of redshift $\ge 3.8$
  quasars from the Palomar Sky Survey (PSS) sample, using the
  Max-Planck Millimetre Bolometer (MAMBO) array at the IRAM 30-metre
  telescope.  Eighteen sources were detected and upper limits were
  obtained for 44 with 3$\,\sigma$ flux density limits in the range
  1.5--4 mJy.  Adopting typical dust temperatures of
  40--50~K, we derive dust masses of a few $\rm 10^8 \, M_{\odot}$ and
  far-infrared luminosities of order $\rm 10^{13} \, L_{\odot}$.  
  We suggest that a substantial fraction of this luminosity arises 
  from young stars, implying star formation rates approaching 
  $\rm 10^3 \, M_\odot yr^{-1}$ or more.  
  The high millimetre detection rate supports current views on a connection
  between AGN and star forming activity, suggesting a parallel
  evolution of the central black hole and of the stellar core of a
  galaxy, although their growth-rate ratio seems higher than the mass ratio
  observed in nearby galaxies. The observed, exceptionally bright 
  objects may trace the peaks of the primordial density field, the 
  cores of future giant ellipticals.
  }

\maketitle

\sloppy

\section{Introduction}

Understanding the history of the formation of stars and massive black
holes during the early stages of galaxy evolution is one of the great
challenges in our quest to understand the history of our Universe.
Recent COBE, SCUBA, and MAMBO observations of the far-infrared to
millimetre extragalactic background show that most of the energy generated
in star formation at high redshift is absorbed by dust, which re-emits
the energy at far-infrared (far-IR) wavelengths, a spectral range that
is red-shifted into transparent submillimetre and millimetre  atmospheric
windows (e.g., Gispert et al. 2000; Ivison et al. 2000a; Blain 2001;
Bertoldi et al. 2000a,b and Carilli et al. 2001a).

In the early stages of formation of galaxies, a large fraction of all 
stars apparently formed in strong bursts within highly obscured regions.
At optical and near-IR wavelengths, much of the star formation
activity in such objects remains not directly visible. Thus from the
ground, only (sub)millimetre observations provide a comprehensive
measure of the energy generated in such objects.  Deep blank field
surveys at 850~$\mu$m  with SCUBA at the JCMT, and at 1.2\,mm with
MAMBO at the IRAM 30-m telescope have by now revealed over one hundred
faint (2 to 10~mJy) background sources, the majority of which
remains invisible in deep optical and near-IR images. This makes
conventional, spectroscopic redshift determinations of this population
mostly impossible.

Global estimates, based on photometric redshifts in the visible/IR
(e.g., Ivison et al. 2000a; Smail et al. 2001) or (sub)millimetre/radio
(Carilli \& Yun 1999), on the far-IR/submillimetre background
intensity and on theoretical modelling (Dole et al. 2000 and references
therein), show that the peak of star formation in the Early Universe
probably occured in the range $z \sim 1-3$, if we assume that much of
the far-IR luminosity arises from star formation.  It is unclear  though,
to what extent nucleus activity contributes to the luminosity of
high-redshift starburst galaxies, and thereby to the far-IR
background. The two brightest objects found in SCUBA and MAMBO blank
field surveys (Knudsen et al. 2001; Bertoldi et al. 2000b)
are in fact quasars, and two of the three SCUBA sources for
which optical emission lines were detected show AGN signatures
(Lilly et al. 1999; Smail et al. 1999; Ivison et al. 2000a)

To understand the relation between the formation of black holes, and  the
formation of galaxies and their bulges, it is necessary to study (aside
from the (sub)millimetre background sources) the emission properties of distant
objects identified at X-ray, optical, near- and mid-infrared, and radio
wavelengths. At very high redshifts ($z>4$), where no blank field millimetre/submillimetre
source has yet been spectroscopically identified, the relation between star
formation and AGN is best studied through targeted (sub)millimetre 
observations of optically selected luminous QSOs and of radio galaxies. 
It is still unclear to what extent the thermal emission of radio-quiet QSOs is
powered by starbursts, or by the black hole accretion.  Just as for
local ULIRGs and Seyfert galaxies (Genzel et al. 1998), there is
increasing evidence that in high-redshift AGN a substantial part of the
thermal emission comes from starbursts.

Well before any SCUBA or MAMBO deep imaging surveys, we were able to
detect seven powerful high-redshift millimetre sources through targeted
observations of ultra-luminous $z>4$ quasars using the 7- and 19-element
bolometer arrays at the IRAM 30-metre telescope (McMahon et
al. 1994; Omont et al. 1996a) and the JCMT (Isaak et al. 1994).
Evidence for the starbust origin of their millimetre emission
arose from the determination of their spectral energy distribution,
including 850--450~$\rm \mu m$ observations
at the JCMT (Isaak et al. 1994, McMahon et al. 1999) and
350~$\mu$m measurements at the CSO (Benford et al.  1999), and from the
detection of CO emission in three objects (Ohta at al. 1996; Omont et
al. 1996b; Guilloteau et al. 1997, 1999).

We here report on a substantial extension of the surveys for 1.2~mm
continuum emission from $z > 4$ ultra-luminous quasars with 18 detections,
which  multiplies by three the number of detections.
Results of 850~$\mu$m observations of a part of the present sample
are presented in Isaak et al. (2001).


        \begin{figure}[hb!]
        \psfig{figure={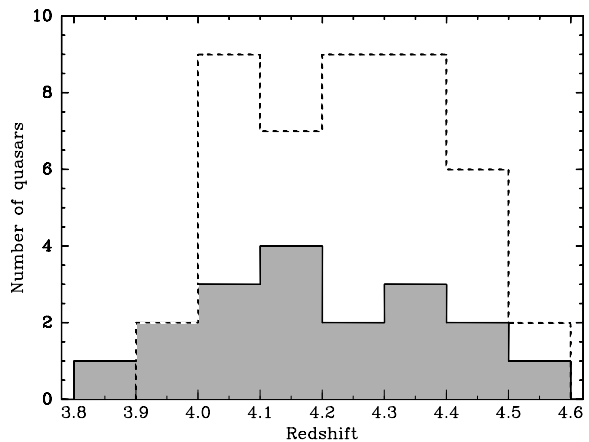},width=9cm,angle=-90}
         \caption{Redshift distribution for the PSS quasars which were
           detected at 1.2~mm (filled histogram) and for those not detected
           (dotted line).}
                \label{figure1}
        \end{figure}

\section{Observations and Results}

The observations were made in February and March 2000 with the
30-metre IRAM telescope at Pico Veleta (Spain), using the {\it Max-Planck
Millimetre Bolometer} (MAMBO; Kreysa et al. 1999). MAMBO is a
37-element bolometer array operating at 300$\,$mK. It is sensitive
between 190 and 315~GHz, with half-power sensitivity limits at 210 and
290~GHz, and an effective bandwidth center for steep spectra at
250~GHz. The bolometer feed horns are matched to the telescope FHWM
beam at 1.2~mm of $10\farcs6$, and they are arranged in an
hexagonal pattern with a beam separation of 22$^{\prime\prime}$. The
sources were observed with the array's central channel, using the
standard on-off observing mode with the telescope secondary chopping
in azimuth by 50$^{\prime\prime}$ at a rate of 2$\,$Hz.  The pointing
was monitored regularly on nearby sources and was found to be stable
typically within $\sim 2^{\prime\prime}$.
Gain calibration was performed using
observations of the planets Mars and Uranus as well as the asteroid
Ceres. A calibration factor of 12500 counts per Jansky is adopted,
which we estimate to be accurate to within 20\%.  The data were
reduced using the MOPSI software package (Zylka 1998).  The
point-source sensitivity of MAMBO at 1.2~mm is 20 to 40
mJy$/\sqrt{\rm Hz}$, depending on weather conditions and the effectiveness
of sky-noise subtraction. In order to remove the strong sky signal from the underlying
source signal, we subtract from each channel the weighted, average
correlated signal from its surrounding channels. 
Subtracting correlated noise has shown to effectively
reduce the sky-noise by typical factors of 2 to 3. Because 
the channels are separated on the array by two beams, point sources
do not produce correlated signals between neighboring channels.
Since the source
signal is not affected by the sky-noise subtraction (which is 
confirmed by extensive tests), 
the error on the final source signal is given
by the remaining noise level of the source channel. Systematic errors
however are introduced by bad pointing, focus, and gain variations that
lead to uncertainties in the flux calibration. 
We have eliminated observations where we
judge such errors to be significantly larger than the errors due
to the normal system- and sky-noise. The remaining flux calibration
uncertainty are of order 10\%.


\begin{table*}[tbhp]
\caption{Quasars detected at 1.2~mm.    
         \label{detections}} 
\begin{center}
\begin{tabular}{lllrrccrl}        
\hline \\[-0.2cm]
\multicolumn{1}{c}{Source} & \multicolumn{1}{c}{$z$} & \multicolumn{1}{c}{$M_B$$^\dag$} 
        & \multicolumn{1}{c}{R.A.} & \multicolumn{1}{c}{Dec.} 
        & \multicolumn{1}{c}{Flux Density} & \multicolumn{1}{c}{Quality$^{\ddag}$} 
	& \multicolumn{1}{c}{Time} & \multicolumn{1}{c}{Comments$^\sharp$} \\ 
  & & & \multicolumn{2}{c}{(J2000.0)} &  \multicolumn{1}{c}{(mJy, $\pm1\,\sigma$)} & & \multicolumn{1}{c}{(sec)} & \\ 
\hline \\[-0.2cm]
 PSS J0209+0517   & 4.18 & $-$28.1 & 02 09 44.7 &    05 17 13.3     &  3.3$\pm$0.6 & A & 2240 &                               \\ [0.1cm]
PSS J0439$-$0207 & 4.40 & $-$27.5 & 04 39 23.1 & $-$02 07 01.7    &  2.3$\pm$0.7 & A & 1940 &   RL$^3$                            \\ [0.1cm]
PSS J0808+5215   & 4.44 & $-$28.7 & 08 08 49.4 &    52 15 15.2    &  6.6$\pm$0.6 & A & 1420 &  RQ$^2$ smm                             \\ [0.1cm]
PSS J1048+4407   & 4.40 & $-$27.4 & 10 48 46.6 &    44 07 12.7    &  4.6$\pm$0.4 & A & 3100 & {\it (b) (c)} W As Comp smm       \\ [0.1cm]
PSS J1057+4555   & 4.12 & $-$28.8 & 10 57 56.3 &    45 55 53.3    &  4.9$\pm$0.7 & A & 1930 & {\it (c)} W As Comp smm         \\ [0.1cm]
BR B1117$-$1329  & 3.96 & $-$28.2 & 11 20 10.2 & $-$13 46 26.2    &  4.1$\pm$0.7 & A & 1420 & {\it (a) (b)} BAL smm1             \\ [0.1cm]
BR B1144$-$0723  & 4.15 & $-$27.7 & 11 46 35.6 & $-$07 40 05.5    &  6.0$\pm$0.7 & A & 2000 & {\it (a)} {\it (b)} BAL smm1     \\ [0.1cm]
PSS J1226+0950   & 4.34 & $-$27.4 & 12 26 23.8 &    09 50 04.7    &  2.8$\pm$0.7 & A & 1410 &                               \\ [0.1cm]
PSS J1248+3110   & 4.35 & $-$27.6 & 12 48 20.2 &    31 10 44.3    &  6.3$\pm$0.8 & A & 2050 &   RQ$^1$ smm                           \\ [0.1cm]
PSS J1253-0228   & 4.00 & $-$27.2 & 12 53 36.4 & $-$02 28 08.3    &  5.5$\pm$0.8 & A & 1290 & RQ$^1$ Comp                          \\ [0.1cm]
PSS J1317+3531   & 4.36 & $-$27.5 & 13 17 43.1 &    35 31 32.3    &  3.7$\pm$1.1 & B & 1160 & RQ$^1$ RQ$^2$ W As                      \\ [0.1cm]
PSS J1347+4956   & 4.56 & $-$28.3 & 13 47 43.3 &    49 56 21.3    &  5.7$\pm$0.7 & A & 1650 &                               \\ [0.1cm]
PSS J1403+4126   & 3.85 & $-$26.8 & 14 03 55.7 &    41 26 16.2    &  1.5$\pm$0.5 & A & 2300 &                               \\ [0.1cm]
PSS J1418+4449   & 4.32 & $-$28.6 & 14 18 31.7 &    44 49 37.6    &  6.3$\pm$0.7 & A & 1290 &  RQ$^1$ smm                             \\ [0.1cm]
PSS J1535+2943   & 3.99 & $-$27.1 & 15 35 53.9 &    29 43 13.0    &  1.9$\pm$0.6 & A & 1410 &            	                  \\ [0.1cm]
PSS J1554+1835   & 3.99 & $-$26.6 & 15 54 09.9 &    18 35 51.0    &  6.7$\pm$1.1 & A & 1170 &  RQ$^1$                             \\ [0.1cm]
PSS J1555+2003   & 4.22 & $-$27.4 & 15 55 02.6 &    20 03 24.4    &  3.1$\pm$0.6 & A & 1410 &                               \\ [0.1cm]
PSS J1646+5514   & 4.10 & $-$28.7 & 16 46 56.5 &    55 14 46.0    &  4.6$\pm$1.5 & B &  380 & RQ$^2$ smm                            \\ [0.1cm]
PSS J1745+6846   & 4.13 & $-$27.0 & 17 45 50.4 &    68 46 20.5    &  2.5$\pm$0.7 & B & 1530 &                               \\ [0.1cm]
PSS J1802+5616   & 4.18 & $-$26.9 & 18 02 48.8 &    56 16 49.6    &  2.8$\pm$0.9 & A & 1540 &                               \\ [0.1cm] 
PSS J2322+1944   & 4.11 & $-$28.1 & 23 22 07.2 &    19 44 23.0    &  9.6$\pm$0.5 & A & 1930 &   RQ$^1$ smm                            \\ [0.1cm]
\hline \\[-0.2cm]
\end{tabular} 
 \end{center}
Notes: Redshifts and coordinates of {\it observed} positions for the PSS quasars are from G. Djorgovski's web page (see text). 
Redshifts and coordinates for the APM BRI quasars are from Storrie-Lombardi et al. (1996). Note that
new astrometry has led to a redetermination of the position of PSS J1057+4555, which is now listed
as 10 57 56.39, +45 55 51.97. For PSS J1048+4407, Isaak et al. (2001) point out that the correct position is 10 48 46.64 +44 07 10.9.

$^{\dag}$ Absolute B-band magnitudes have been calculated as described in Isaak
et al. (2001), for $H_0 = 50$~$\rm km\,s^{-1}$ Mpc$^{-1}$ and 
$q_0$=0.5. 

$^{\ddag}$ Quality of the observations: A is good and B is poor. 

$^\sharp$ {\it (a)} Omont et al. (1996a) {\it (b)} Guilloteau et
al. (1999)  {\it (c)} see text for radio properties.  RQ=radio-quiet,
RL=radio-loud: the subscripts 1, 2 and 3 refer to Carilli et al. (2001b),
Stern et al. (2000) and Isaak et al. (2001), respectively; BAL=Broad Absorption Lines;
W=Weak (and broad) emission lines; As=Associated absorption system;
Comp=companion(s); smm and smm1 refer to 850~$\mu$m detections reported by
Isaak et al. (2001) and McMahon et al. (1999), respectively.
 \end{table*}

Most of our sources were observed several times on different dates, and
we were able to check for consistency in the resulting fluxes. We
noticed that several observing periods yielded source fluxes
which were altogether inconsistent with observations of the same objects
on other dates. We suspect that during such periods, either the
telescope pointing, optics, or an unstable atmosphere lead to a
significant sensitivity loss, and we therefore ignored observations
during such suspect periods.  For a reliable flux determination, a
source should be observed at least on two
different dates and yield consistent results, or a reliable, nearby flux
calibrator should have been observed along with the source.
These considerations were used to define a quality factor
which we list with the fluxes in Table 1.

The high-redshift quasars were selected from the multicolor Palomar
Digital Sky Survey (DPOSS: Kennefick et al. 1995a and b; Djorgovski et
al. 1999), that were available at the time from G.~Djorgovski's web
page: 

http://astro.caltech.edu/~george/z4.qsos

In order to study the most luminous objects, we arbitrarily
limited the observations to sources brighter
than $\rm M_B = -27.0$ with a few exceptions.  Altogether, 62 PSS sources were observed out
of the 90 PSS quasars listed by G.~Djorgovski at the time.

\begin{figure}[ht!]
        \psfig{figure={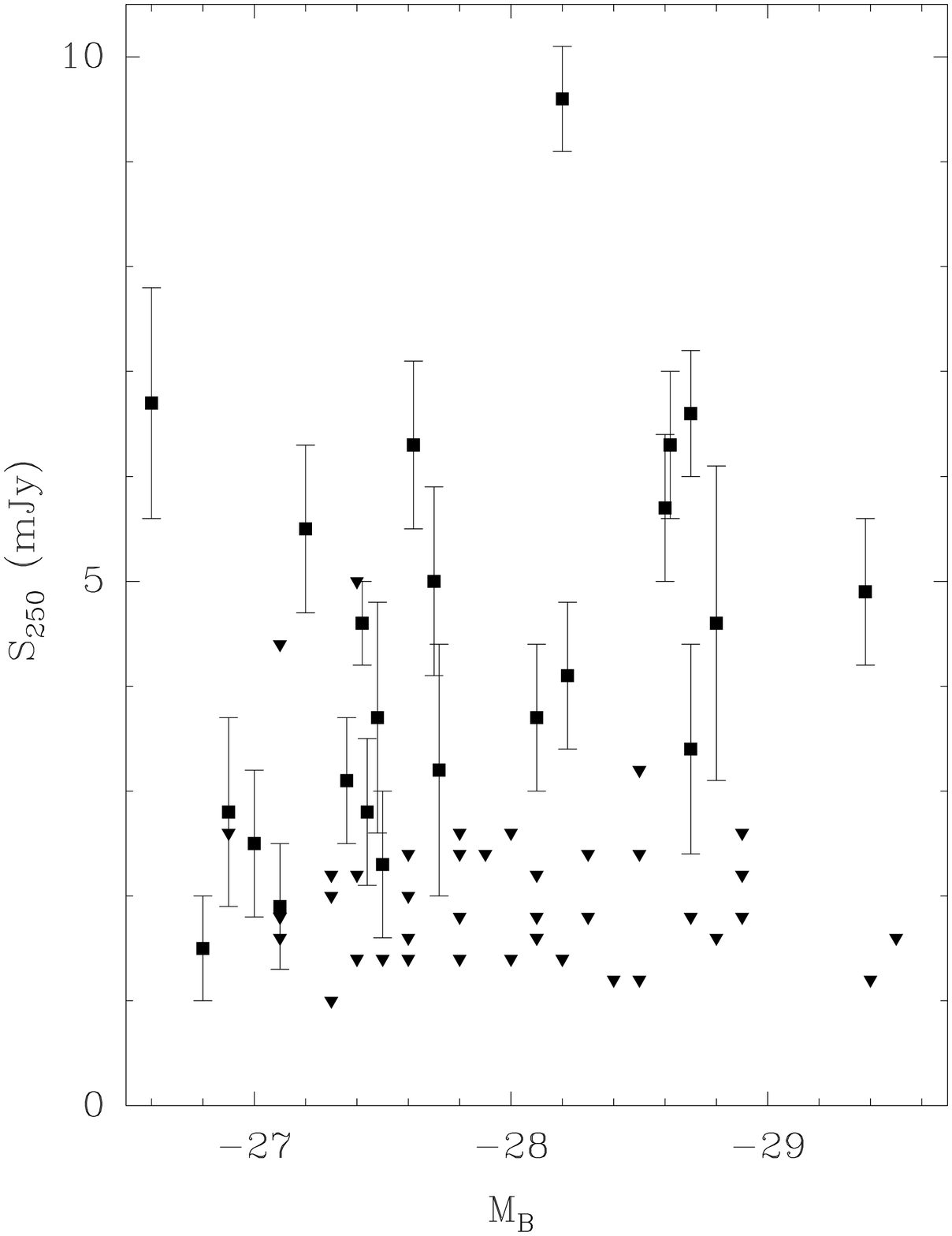},width=8cm}
	\caption{Observed 1.2~mm flux density versus $\rm M_B$, the
                 optical absolute magnitude in the B rest-frame band.
	         QSOs detected at 1.2~mm (Table~1) are
                 shown as filled squares. The triangles represent
		 $2\,\sigma$ upper limits for sources not detected
                 at 1.2~mm (Table~2) - plotted are the $2\,\sigma$ values or
                 the flux density plus $\sigma$ if $S_{250} > 2\,\sigma$.}
 	\label{figure2}
        \end{figure}

Eighteen new sources were detected at levels $\ge 3 \,\sigma$,
of which 12 sources were detected at levels $\ga 4 \,\sigma$, as outlined in
Table~1. Figure~\ref{figure1} shows that their redshifts are
uniformly distributed between $z = 3.9$ and 4.5.
Figure~\ref{figure2} displays the observed 1.2~mm flux density $S_{250}$ versus $\rm M_B$,
the optical absolute magnitude in the B rest-frame band, which represents
well the bolometric magnitude, $L_{\rm bol}$ (Sec.~3.6). It shows that
there is no strong correlation between $\rm M_B$  and $S_{250}$.

Nine of the quasars are detected with a 1.2~mm flux density $\ga 5 \, \rm mJy$.
PSS J2322+1944, with a flux density of 9.6$\pm$0.6~mJy, is the strongest millimetre
source at $z \ge 4$ just after BR1202$-$0725, which has a 1.2~mm
flux density of 18.5$\pm$2~mJy (derived from Guilloteau et al. 1999).
Both sources are exceptional and are
indeed brighter than any other high-redshift source known, except the
most strongly lensed sources, H1413+117 (Cloverleaf), FIRAS10214+4724 and 
APM08279+5255 (however, see also the case of BRI B1335$-$0417, Guilloteau et al. 1999).

\begin{table*}[tbhp]
\caption{Quasars with upper limits at 1.2~mm.
         \label{nondetections}}
\begin{center}
\begin{tabular}{lllrrrcrl}
\hline \\[-0.2cm]
\multicolumn{1}{c}{Source} & \multicolumn{1}{c}{$z$} & \multicolumn{1}{c}{$M_B$$^\dag$}
        & \multicolumn{1}{c}{R.A.} & \multicolumn{1}{c}{Dec.}
        & \multicolumn{1}{c}{Flux Density} & \multicolumn{1}{c}{Quality $^{\ddag}$} & \multicolumn{1}{c}{Time} & \multicolumn{1}{c}{Comments}  \\
  & & & \multicolumn{2}{c}{(J2000.0)} & \multicolumn{1}{c}{(mJy, $\pm1\,\sigma$)} & & \multicolumn{1}{c}{(sec)}  \\
\hline \\[-0.2cm]
PSS J0014+3032   & 4.40 & $-$28.1 & 00 14 43.0  &    30 32 03.3  &    $-0.5\pm0.9$ & B & 780 &               \\ [0.1cm]
PSS J0121+0347   & 4.13 & $-$27.6 & 01 21 26.1  &    03 47 07.2  &    $1.7\pm1.2$  & B &  780 & RL$^2$, RL$^3$           \\ [0.1cm]
PSS J0121+3453   & 4.22 & $-$27.6 & 01 21 11.5  &    34 53 00.0  &    $0.7\pm1.0$  & B &  780 &              \\ [0.1cm]
PSS J0131+0633   & 4.43 & $-$27.8 & 01 31 12.2  &    06 33 40.0  &    $1.3\pm1.2$  & B &  780 &              \\ [0.1cm]
PSS J0134+3307   & 4.52 & $-$27.4 & 01 34 21.5  &    33 07 55.4  &    $3.5\pm1.5$  & A &  780 & RQ$^2$           \\ [0.1cm]
PSS J0211+1107   & 3.99 & $-$27.1 & 02 11 20.1  &    11 07 16.0  &    $-1.0\pm0.9$ & A & 1420 &  RL$^3$            \\ [0.1cm]
PSS J0452+0355   & 4.38 & $-$27.6 & 04 52 51.5  &    03 55 57.0  &    $0.2\pm0.8$  & A & 1300 & smm             \\ [0.1cm]
PSS J0747+4434   & 4.42 & $-$28.5 & 07 47 49.7  &    44 34 16.0  &    $0.1\pm1.2$  & A &  780 & RQ$^2$           \\ [0.1cm]
PSS J0852+5045   & 4.20 & $-$28.2 & 08 52 27.3  &    50 45 10.9  &    $0.6\pm0.7$  & A & 1300 &              \\ [0.1cm]
PSS J0926+3055   & 4.19 & $-$29.5 & 09 26 36.3  &    30 55 04.9  &    $1.0\pm0.8$  & A & 2330 &              \\ [0.1cm]
PSS J0950+5801   & 3.97 & $-$28.3 & 09 50 13.9  &    58 01 38.0  &    $0.4\pm0.9$  & A &  780 &              \\ [0.1cm]
PSS J0955+5940   & 4.34 & $-$28.9 & 09 55 11.3  &    59 40 30.5  &    $-1.3\pm1.1$ & A &  900 &              \\ [0.1cm]
PSS J0957+3308   & 4.25 & $-$28.3 & 09 57 44.5  &    33 08 20.4  &    $0.1\pm1.2$  & B &  640 &              \\ [0.1cm]
PSS J1026+3828   & 4.18 & $-$27.8 & 10 26 56.7  &    38 28 44.9  &    $0.4\pm0.7$  & B &  900 &              \\ [0.1cm]
PSS J1058+1245   & 4.33 & $-$28.8 & 10 58 58.4  &    12 45 54.8  &    $0.1\pm0.9$  & B & 1290 &              \\ [0.1cm]
PSS J1159+1337   & 4.08 & $-$28.8 & 11 59 06.6  &    13 37 37.6  &    $0.9\pm0.8$  & B & 1380 & RQ$^2$           \\ [0.1cm]
PSS J1315+2924   & 4.18 & $-$27.4 & 13 15 39.8  &    29 24 39.1  &    $0.7\pm0.7$  & A &  900 &              \\ [0.1cm]
PSS J1326+0743   & 4.17 & $-$28.4 & 13 26 11.8  &    07 43 57.5  &    $-0.3\pm0.6$ & A & 1800 &              \\ [0.1cm]
PSS J1339+5154   & 4.08 & $-$27.2 & 13 39 13.0  &    51 54 03.8  &    $1.4\pm0.6$  & B & 1540 &              \\ [0.1cm]
PSS J1401+4111   & 4.01 & $-$27.9 & 14 01 32.8  &    41 11 50.9  &    $-0.4\pm1.2$ & B &  520 &              \\ [0.1cm]
PSS J1430+2828   & 4.30 & $-$27.1 & 14 30 31.6  &    28 28 34.3  &    $3.2\pm1.2$  & B &  650 &              \\ [0.1cm]
PSS J1432+3940   & 4.28 & $-$28.2 & 14 32 24.8  &    39 40 24.4  &    $-0.1\pm0.7$ & A & 1420 &              \\ [0.1cm]
PSS J1435+3057   & 4.35 & $-$28.5 & 14 35 23.5  &    30 57 22.1  &    $2.4\pm1.6$  & A &  390 & RQ$^2$ W         \\ [0.1cm]
PSS J1443+2724   & 4.42 & $-$27.3 & 14 43 31.2  &    27 24 36.9  &    $-0.1\pm0.5$ & A & 2190 & RQ$^2$           \\ [0.1cm]
PSS J1443+5856   & 4.27 & $-$28.5 & 14 43 40.7  &    58 56 53.3  &    $0.7\pm0.6$  & B & 2320 &              \\ [0.1cm]
PSS J1500+5829   & 4.22 & $-$27.8 & 15 00 07.7  &    58 29 37.7  &    $0.4\pm0.9$  & A &  640 &              \\ [0.1cm]
PSS J1506+5220   & 4.18 & $-$27.7 & 15 06 54.6 &    52 20 04.6    &  $3.2\pm1.2$ & A &  780 &                               \\ [0.1cm]
PSS J1531+4517   & 4.20 & $-$27.4 & 15 31 29.9  &    45 17 07.9  &    $0.9\pm0.7$  & B & 1940 &              \\ [0.1cm]
PSS J1543+3417   & 4.39 & $-$29.4 & 15 43 40.4  &    34 17 44.5  &    $0.3\pm0.6$  & A & 1290 &              \\ [0.1cm]
PSS J1615+1803   & 4.01 & $-$27.6 & 16 15 22.9  &    18 03 56.0  &    $-0.1\pm0.7$ & A & 1420 &              \\ [0.1cm]
PSS J1618+4125   & 4.21 & $-$27.4 & 16 18 22.7  &    41 25 59.7  &    $1.2\pm1.1$  & B &  500 & RQ$^2$           \\ [0.1cm]
PSS J1633+1411   & 4.35 & $-$27.8 & 16 33 19.6  &    14 11 42.3  &    $1.1\pm1.2$  & B &  640 &              \\ [0.1cm]
PSS J1721+3256   & 4.03 & $-$27.1 & 17 21 06.7  &    32 56 35.9  &    $-0.8\pm0.8$ & A & 1160 & RQ$^2$ Comp      \\ [0.1cm]
PSS J2134+0817   & 4.** & $-$27.1 & 21 34 43.2  &    08 17 28.6  &    $0.3\pm0.9$  & A &  780 &              \\ [0.1cm]
PSS J2154+0335   & 4.46 & $-$27.5 & 21 54 06.7  &    03 35 39.2  &    $0.4\pm0.7$  & A & 1530 &              \\ [0.1cm]
PSS J2155+1358   & 4.26 & $-$28.1 & 21 55 02.2  &    13 58 25.5  &    $0.6\pm0.8$  & B &  910 &              \\ [0.1cm]
PSS J2203+1824   & 4.38 & $-$28.7 & 22 03 43.4  &    18 28 13.1  &    $0.4\pm0.9$  & B &  780 &              \\ [0.1cm]
PSS J2238+2603   & 4.03 & $-$28.9 & 22 38 41.6  &    26 03 45.3  &    $2.5\pm1.3$  & B &  750 &              \\ [0.1cm]
PSS J2241+1352   & 4.46 & $-$28.0 & 22 41 47.8  &    13 52 02.0  &    $-0.0\pm1.3$ & A &  510 &              \\ [0.1cm]
PSS J2244+1005   & 4.04 & $-$26.9 & 22 44 05.4  &    10 47 38.6  &    $-1.5\pm1.3$ & A &  520 &              \\ [0.1cm]
PSS J2256+3230   & 4.04 & $-$26.0 & 22 56 10.4  &    32 30 18.5  &    $-1.7\pm0.8$ & B &  780 &  RL$^3$            \\ [0.1cm]
PSS J2315+0921   & 4.52 & $-$28.1 & 23 15 59.1  &    09 21 43.6  &    $-0.4\pm1.1$ & B &  910 &              \\ [0.1cm]
PSS J2323+2758   & 4.18 & $-$27.3 & 23 23 40.9  &    27 58 00.4  &    $1.7\pm1.1$  & A &  780 &              \\ [0.1cm]
PSS J2344+0342   & 4.30 & $-$28.1 & 23 44 03.2  &    03 42 26.4  &    $0.1\pm0.8$  & A & 1540 &              \\ [0.1cm]
\hline \\[-0.2cm]
\end{tabular}
\end{center}
\hspace{1cm} Note: Same notations as in Table 1.  
\end{table*}

In addition, three quasars (PSS J1048+4407, BR B1117$-$1329 and
BR B1144$-$0723) which were previously detected at the 30-metre
telescope, but were not detected at similar levels in observations with
the Plateau de Bure interferometer (Omont et al. 1996a, Guilloteau et al. 1999 Table 2), were
remeasured.  Their new 1.2~mm flux densities are between 4 and 5~mJy
(Table~1), confirming the previous 30-metre data ($3 \pm 1$, $4.09 \pm
0.81$, and $\rm 5.85 \pm 1.03 \, mJy$, respectively, at 1.25~mm).
These values are also compatible with the 850~$\mu$m flux densities
measured with SCUBA at the JCMT: $\rm 12 \pm 3 \, mJy$ for PSS
J1048+4407 (Isaak et al. 2001), $\rm 13 \pm 1 \, mJy$ and $\rm 7 \pm 2
\, mJy$ for BR B1117$-$1329 and BR B1144$-$0723, respectively (McMahon
et al. 1999). This consistent set of data makes the flux densities
measured at 1.2~mm and 850~$\mu$m with single dishes very robust for
the three sources.  For BR B1144$-$0723 a new measurement at
850~$\mu$m with a better sensitivity would be desirable. As a consistency
check, this source was remeasured in February 2001 with the 30-metre telescope
at 1.2~mm yielding a result which is perfectly consistent with the
measurements done in 2000. The value of the flux density listed in
Table~1 includes this recent measurement.

Interestingly, all these single-dish measurements (except at
850~$\mu$m for BR B1144$-$072) are incompatible, for point sources,
with the results of the IRAM interferometer measurements at 1.35~mm
reported by Guilloteau et al. (1999).  If PSS J1048+4407, BR
B1117$-$1329 and BR B1144$-$0723 are unresolved by the interferometer
beam ($\approx 3^{\prime\prime} \times 2^{\prime\prime}$), their
1.35~mm flux densities should be limited to $\rm 0.25 \pm 0.68 \,
mJy$, $\rm 0.64 \pm 1.47 \, mJy$ and $\rm 0.54 \pm 0.78 \, mJy$,
respectively. The most natural explanation of the poor agreement
between the single dish and the interferometer would be that the dust
millimetre emission of these three sources is extended with respect to
the interferometer beam. Such extensions have been reported for other
high redshift sources (Omont et al. 1996b; Papadopoulos et al. 2000,
2001; Richards 2000). However, this needs confirmation for PSS
J1048+4407, BR B1117$-$1329 and BR B1144$-$0723.

Forty-four PSS quasars were not 3$\,\sigma$ detected at 1.2~mm, with
typical 3$\,\sigma$ flux density upper limit of 1.5--4~mJy. Table~2
lists their names, redshifts, and the 1.2~mm flux densities with
$\pm1\,\sigma$ errors.

We have detected 30\% of the 62 new sources observed. Such a detection
rate is consistent with previous studies, in particular with that of
Omont et al. (1996a) on a three times smaller sample of similar
brightness APM BRI survey quasars. It is also consistent with the
parallel, deeper MAMBO study of a sample of 41 SDSS quasars (Carilli
et al. 2000, 2001a).  It should be stressed that such bright $z > 4$
quasars with M$_B$ $< -27$ are rare, with an average surface density
of one per 50-100~deg$^2$ (Storrie--Lombardi et al. 2001).  The number
of $z>4$ quasars detectable in the (sub)millimetre with flux densities
$\rm \ga 5 \, mJy$ is thus less than one per 100~deg$^2$.

\section{Discussion}

\subsection{The case for thermal dust emission}

There is little doubt that, for nearly all the high-redshift quasars
detected in the millimetre, this emission is due to dust, and not to
synchrotron radiation. This is shown either by their large and positive
submillimetre/millimetre spectral index, which is characteristic of
dust emission, or by the fact that they are not radio-loud at
1.4~GHz or 5~GHz. Both indicators have well established the dust origin of the
millimetre emission of the $z > 4$ quasars detected until now at
1.2~mm (Isaak et al. 1994; Benford et al. 1999; McMahon et al. 1999;
Carilli et al. 1999; Kawabe et al. 1999; Priddey and McMahon 2001).

The characteristics of the dust emission of previously detected
sources have been derived by combining observations made at various
(sub)millimetre wavelengths. For instance, by comparing 350~$\mu$m
measurements of six high-$z$ quasars with 1.3~mm - 450~$\rm \mu m$
flux densities, Benford et al. (1999) showed that most of the dust
contributing to the millimetre/submillimetre emission is at a
temperature close to 50~K, with a power-law emissivity index
$\beta$=1.5$\pm$0.2.  Recently, Priddey \& McMahon (2001) compiled an
average spectral energy distribution (SED) for the $z > 4$ quasars
detected by Omont et al.  (1996a) and derived $T = 41 \pm 5 \, \rm K$
and $\beta$=1.95$\pm$0.3. Note that these data are not exclusive of
dust at higher temperature, as seen in H1413+117 (Cloverleaf),
F10214+4724 or APM08279+5255, and expected from models (e.g., Andreani
et al. 1999). The luminosities which have been derived in previous
studies from the SEDs of $z > 4$ quasars detected in the millimetre
are $\ge 10^{13} ~\rm L_{\odot}$ (see also Eq.~3),
slightly greater than the typical luminosity of ultra-luminous
infrared galaxies such as Arp~220.

Similarly, radio and submillimetre data begin to be available for a
part of the broader PSS sample, especially for the sources that we
have detected at 1.2~mm. All results point to dust emission. A very
deep radio VLA study of nine PSS QSOs, among the strongest 1.2~mm
sources of Table 1, shows that their millimetre emission is most
likely thermal dust emission (Carilli et al. 2001b). Six of them are
radio-quiet (Table 1).
PSS J1057+4555 is shown as a weak radio-loud quasar and PSS J1048+4407 as a weaker
one. However, the steep radio spectrum of both sources makes it
unlikely that their large 1.2~mm flux densities result from
synchrotron emission.

Stern et al. (2000) and Isaak et al. (2001) have
also addressed the radio properties
of optically selected $z > 4$ quasars.
From a deep 5~GHz imaging survey, Stern et al. have identified as radio-quiet
two of the sources of Table 1 (detected at 1.2~mm) and seven of Table 2 (undetected at 1.2~mm).
From a comparison with the shallow  NVSS 1.4~GHz survey, Isaak et al. (2001) have
identified as radio-loud one source detected at 1.2~mm, PSS J0439$-$0207, and
three undetected ones (Table 2).

Seven of the 1.2~mm MAMBO PSS detections (marked by `smm' in
Table~1) were also detected at 850~$\rm \mu m$ by Isaak et al. (2001),
indicating 850~$\rm \mu m$/1.2~mm flux ratios which are characteristic
of dust emission, in agreement with their radio properties. A detailed
discussion of the submillimetre properties of these quasars and
parameters inferred from them is left to a later publication (Priddey
et al., in preparation).

\subsection{Dust mass}

In the likely case that the millimetre emission is optically thin, the
total mass of dust is directly proportional to the millimetre flux
        \begin{equation}
        M_{d} = {S_{250} \, D^2_L\, \over (1+z) \,
        \kappa_d(\nu_r) \, B(\nu_r,T_d)}~,
        \end{equation}
where $S_{250}$ is the flux density at 1.2~mm (i.e., 250~GHz),
$\nu_r$ is the rest-frame frequency, $\kappa_d(\nu_r)$
is the dust absorption coefficient, $B(\nu_r,T_d)$ is the Planck
blackbody function and $D_L$ is the luminosity distance which depends
on the cosmological parameters (see, e.g., Eq.~(1) in McMahon et
al. 1994).  Hereafter, we adopt a standard Einstein-de Sitter
cosmology with $H_0 = 50$~$\rm km\,s^{-1}$ Mpc$^{-1}$ and
$q_0$=0.5. To convert to a Lambda cosmology with $\Omega _M=0.3$,
$\Omega _\Lambda =0.7$, and $H_0$=65~$\rm km \, s^{-1}$ Mpc$^{-1}$,
the masses and luminosities given here should be multiplied by
$\approx1.4$. For the median value of the redshift distribution of
Fig. 1, $z\approx 4.2$, the rest-frame wavelength is
$\approx$\,230\,$\mu$m. Equation (1) can then be
rewritten as
        \begin{equation}
        M_{d} \approx 1.1 \times 10^8 \, F(T_d) \,
 \left(7.5~cm^2 g^{-1}\over\kappa_{d}(230\mu m)\right) \,         \left(S_{250}\over\rm mJy\right)\, M_{\odot}
        \end{equation}
where $F(T_d)\,=\,(\exp[h\nu/kT_d]-1)/(\exp[h\nu/k\cdot 50\rm K]-1)$
accounts for the temperature dependence. The dust absoption coefficient at 230~$\mu$m,  
$\kappa_{d}(230\,\mu$m)\,=\,7.5\,$\rm cm^2 g^{-1}$, is the value extrapolated from the value 
at 125~$\mu$m of Hildebrand (1983), $\kappa_d(125\, {\rm
\mu \rm m}) \, = \, 18.75 \, \rm cm^2 g^{-1}$ and a
frequency dependence of $\beta$\,=\,1.5. 
Using Eq.~(2) and $T_d \sim 50 \, \rm ~K$, the mass of dust of the
sources detected at 1.2~mm (Table~1) is in the range $2 \times 10^8$
to $\rm 10^9 \, (7.5/\kappa_{d230}) \, M_\odot$ with a median value of $\rm \sim 5 \times
10^8 \, (7.5/\kappa_{d230}) \, M_\odot$.

Equation (2) is consistent with previous formulae (e.g., Omont et
al. 1996a; Rowan-Robinson 2000).  However, note that for
$T_d\,\sim40-50 \rm \,K$ the above relation is slightly different from
Eq.~(1) in Omont et al. (1996a) where a dust temperature of 80~K was
assumed.  For $T_d\sim50$~K, the temperature dependence of $M_d$ is
roughly $T_{d}^{-1.6}$.

Note that other temperature dust components may be present. As quoted
in Sec. 3.1, in the few quasars for which higher frequency
measurements are available, a warm dust component is evident. A
significant cold dust component could also be hidden below the
spectral energy distribution of the warm dust.  In the most extreme
case of dust at the CMB temperature at $z=4$, $T_d\approx 13$~K, a
given flux density at 1.2~mm could arise from five times the amount of
dust we would estimate assuming $T_d=50$~K. More realistically, we
estimate that within the flux density uncertainties a hidden cold dust
component could still be present with perhaps about twice the mass we
estimate ignoring such a cold component.

We should emphasize that the estimates of M$_d$ 
depend on the assumed values of the dust temperature and emissivity. 
The latter remains highly uncertain, particularly for the unknown properties of dust in luminous $z>4$ quasars. 
Dust masses $M_{d} \approx \, 1-4 \, \times \, 10^7 \, M_{\odot}$ are 
derived by Andreani et al. (1999) in their Table 3 for the sources
whose 1.25~mm flux values are reported by Omont et al. (1996a). The
origin of such low values of $M_{d}$ is probably due to the high dust temperature
resulting from AGN heating and to the adopted grain
model (Granato \& Danese 1994; Rowan-Robinson 1986) which has a large
emissivity at millimetre wavelengths.

\subsection{Dust heating: Starburst or AGN?}

As discussed by various authors (Sanders \& Mirabel 1996; Andreani et
al. 1999; Rowan-Robinson 2000; Yun et al. 2000; Carilli et al. 2000),
for ultra luminous infrared galaxies (ULIRGs) it is difficult to
disentangle the dust heating contributions from a starburst and from the
central AGN, even in the local universe (Genzel et al.  1998; Downes \&
Solomon 1998; Veilleux et al. 1999), where such objects are bright  but rare.
Due to the lack of detailed spectral, geometric, and
kinematic information, for sources at very high redshifts only indirect
arguments can be made in favour of one or the other heating mechanism.

An argument that favours AGN as the dominant heating source is the
large power available that could heat the dust even at kpc distances.
Simple torus models can account for the observed SED of AGN at small
and moderate redshift (Sanders et al. 1989; Granato \& Danese 1994;
Granato et al. 1996; Andreani et al. 1999). In particular, AGN heating
seems required to explain the dust detected at higher temperature in
lensed sources such as H1413+117, FIRAS10214+4724 and APM08279+5255,
if not due to differential lensing (e.g., Rowan-Robinson 2000 and
references therein).

Starbursts as a source of the far-IR luminosity are favoured by the large
observed dust masses, $\rm \ga 10^8 \, ~M_\odot$, which imply the presence
of $> \rm 10^{10} \, M_{\odot}$ of gas, much of which may be molecular
as indicated by the detection of CO in three $z > 4$ quasars (Ohta et
al. 1996; Omont et al.  1996b; Guilloteau et al.  1997, 1999). Such gas
masses are typical for local ULIRGs (Solomon et al.  1997; Downes \&
Solomon 1998) and hyper-luminous infrared galaxies (Rowan-Robinson
2000). The high gas masses, the dust temperatures (Benford et al. 1999),
and the faint radio emission (Yun et al. 1999; Kawabe et al. 1999;
Carilli et al.  2000, 2001b) are consistent with starbursts with star
formation rates, $\dot M_{\rm SF}$, of order $\rm 1000~M_\odot yr^{-1}$.
The total far-IR luminosity, $L_{\rm FIR}$, which we take to be
the integrated luminosity at rest wavelengths $> 50 \, \rm \mu m$, has
been well determined from millimetre to 350~$\mu$m detections (Benford
et al. 1999) for several objects at $z>4$.

For a modified black-body with $T_d = 50 \, \rm K$ and $\beta = 1.5$,  
$L_{\rm FIR}$ is related to $M_d$ through
$L_{\rm FIR} = 3.3 \times 10^{12} ~ (M_d/10^8 M_\odot) ~(\kappa_{d230}/7.5\,\rm cm^2 g^{-1}) ~~ \rm L_\odot$
- see Eq.~(5) in McMahon et al. (1999). Therefore, from Eq.~(2)
	
        \begin{equation}
        L_{\rm FIR} \sim  3.6\times 10^{12} ~(S_{250}/
        {\rm mJy} ) ~~ \rm L_\odot ~.
        \end{equation}

\subsection{Starbursts: Implied star formation rates}

\begin{figure}[]
        \psfig{figure={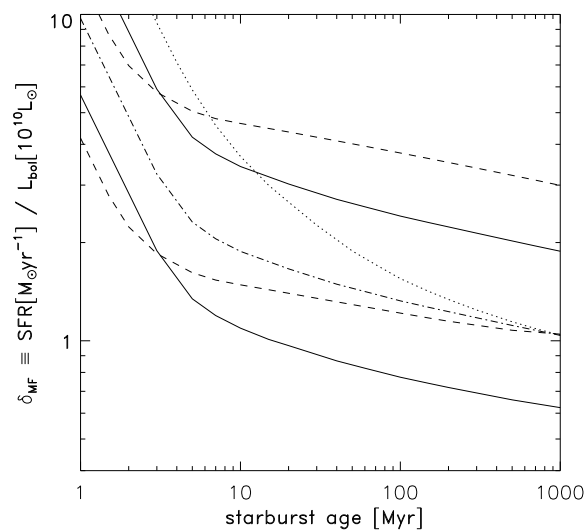},width=8.5cm}
        \caption{Star formation rate per total luminosity,
          $\delta_{\rm MF}$, for continuous starbursts as a function of
          starburst age, adopting different IMFs. {\it Solid lines:}
          Salpeter IMF with $m_l=0.1\rm M_\odot$ (upper solid line) and
          $m_l=1.6\rm M_\odot$ (lower solid line), with stellar
          lifetimes and luminosities taken from Schaerer et al. (1993) 
          evolution tracks.  {\it Dashed lines:} As for solid lines, but adopting
          approximate lifetimes and luminosities from Telesco \& Gatley
          (1984).  {\it Dash-dotted line:} Like upper solid line but
          assuming a flat IMF below $1\rm M_\odot$.  {\it Dotted line:}
          As upper solid line, but assuming a log-normal IMF: $d\log
          N/d\log m = - \log m$.}
\label{figure3}
\end{figure}

Let us consider the case that the far-IR luminosity
arises  exclusively from starbursts, and that the stellar
radiation is completely absorbed and reemitted by warm dust.
Then the far-IR luminosity equals the starburst's total luminosity.
In a continuous starburst the star formation rate relates to the
far-IR luminosity
\begin{equation}
        \dot M_{\rm SF} ~=~\delta_{\rm MF}~
        (L_{\rm FIR}/\rm 10^{10}L_\odot)~~M_\odot yr^{-1}~,
\end{equation}
where $\delta_{\rm MF}$ is a function of the mass function (MF), i.e.,
of the present mass composition of the stellar population.  To
determine $\delta_{\rm MF}$, as an example, we first consider stars
forming with a Salpeter initial mass function (IMF), $d\log N/d\log m
= -2.35$, with a low mass cutoff, $m_l$. We can adopt the approximate
expressions for the stellar luminosities and lifetimes given by
Telesco \& Gatley (1984).  Then for a 100 Myr old continuous
starburst, $\delta_{\rm MF} = 1.2$ and 3.8, for a low mass cutoff
$m_l= 1.6 \rm~ M_\odot$ and $\rm 0.1 \, M_\odot$, respectively,
(Fig.~\ref{figure3}).  These values of $\delta_{\rm MF}$ are lower by
a factor of two from those which are commonly adopted from Thronson \&
Telesco (1986), which were computed for a continuous starburst of age
2 Myr.  Such a short starburst age is certainly inappropriate for the
massive starbursts we observe in high-redshift galaxies, which more
likely occur on timescales comparable to the dynamical timescale of
the starburst regions, $10^7 - 10^8$ yr.

We have alternatively computed $\delta_{\rm MF}$ using recent stellar
evolution models (Schaerer et al. 1993), considering a range of
metallicities and IMFs.  For starburst ages of 10--100~Myr, and a more
realistic flat IMF at low masses ($d\log N/d\log m = -1$ at $m<1\rm \,
M_\odot$, $m_l= 0.1 \, \rm M_\odot$), $\delta_{\rm MF}$ ranges between
0.8 and 2 (Fig.~\ref{figure3}), well in agreement with previous
studies (see, e.g., Kennicutt 1998; Barger et al. 2000; Rowan-Robinson
2000). Whilst the exact value of $\delta_{\rm MF}$   for the $z > 4$ 
sample discussed in this paper is unknown, it is unlikely that it will
deviate much from this range.


In reality, the far-IR luminosity may differ from the starburst's
total luminosity for two main reasons: either part of the stellar
radiation is not absorbed by dust and escapes, or part of the dust
heating is achieved by the quasar radiation. Let $\delta_{\rm SB}~$ be
the ratio between the starburst's total luminosity and $L_{\rm
FIR}$. Note that $\delta_{\rm SB}~$ is just an empirical definition
which can depend on several parameters.  It follows from Eqs.~(3) and
(4) that the star formation rate of a continuous starburst relates to
the 1.2~mm flux density, $S_{250}$,
\begin{equation}
        \dot M_{\rm SF} ~\approx~ 360 ~ \delta_{\rm MF}~ \delta_{\rm SB}~
        (S_{250}/ \rm mJy) ~ M_\odot yr^{-1}~.
\end{equation}

\subsection{Heavy element production and FIR emission}

The dust seen through its far-IR emission was generated by stars that
formed during current or past star formation events.  The mass of dust
may provide an estimate of the minimum amount of heavy elements (i.e.,
$\ge 12$ amu) already synthesized, and hence of the energy radiated
from nucleosynthesis and of the total star formation. The energy
directly generated in the nucleosynthesis of a mass $M_m$ of heavy
elements is $\sim$0.008 $M_m$ c$^2$.
Accounting for the energy released in the formation of helium which is
simultaneously synthesized in massive stars, but not transformed in heavy
elements, one may estimate that the radiative energy, $E_m$, emitted by the stars which
release a mass, $M_m$, of heavy elements into the interstellar medium,
is of the order of
        \begin{equation}
        E_m ~\sim~ 0.02 \, M_m \, c^2.
        \end{equation}
(see Harwit 1999). From Eqs. (4) and (6), the factor of proportionality 
betwen the total mass of formed stars and of released heavy elements is
	 \begin{equation}
        M_{\rm SF}/M_m  \approx 30 ~ \delta_{\rm MF}~.
       \end{equation} 

The mass of heavy elements ever released in a galaxy must be larger 
than the mass of dust we can observe through its FIR emission.
We introduce
\begin{equation}
\delta_{md} ~\equiv~ M_m/M_{d} ~>~1
\end{equation}
as the ratio between the total mass of released heavy elements and the
visible dust mass. This ratio accounts for heavy elements in the gas
phase and in the intergalactic medium, or reprocessed in new stars, and for cold dust.
In the local interstellar medium, the mass fraction of heavy elements
locked in dust grains is $\sim$40\%, when adopting a total abundance
of heavy elements $\sim$2/3 solar (Meyer et al. 1998) and a
gas-to-dust ratio $\sim$190 (Allen 1999, Table 7.12).  This fraction
is unknown in local starburst galaxies, but there are no indications
that it is very different, and the presence of a consistent amount of
gas phase oxygen and carbon at least is attested through the detection
of CO.  Adopting the Galactic dust fraction of heavy elements, we
speculate that in high-redshift starbursts, $\delta_{md}\approx 2.5$,
but that this value is uncertain by a factor of two in either
direction.

Some of the heavy elements produced in a starburst may be blown into
the intergalactic medium or scattered throughout the galaxy to regions
far from radiative energy sources. The dust in this component may not
be heated sufficiently to contribute noticably to the integrated far-IR
emission.  The intergalactic ejection of enriched gas is expected to
be particularly significant in extreme starbursts such as those we
might be observing in the high-redshift quasars, and could well be
significant in the observed enrichment of the intergalactic medium of
galaxy clusters.  The fraction of metal to {\it observed} dust may
thus be significantly larger than the quoted value of 2.5.  In
conclusion, it is difficult to estimate a more precise value for
$\delta_{md}$, which probably ranges between 1 and 10.

If the radiative energy that is emitted along with the generation of
heavy elements in a galaxy is released over a timescale $\tau_{\rm
SF}$, the average total luminosity of the stars during that time
would be
\begin{eqnarray}
&&L_m ~=~ E_m ~/~ \tau_{\rm SF} ~=~ 0.02 ~\delta_{md}~ M_d~ c^2 ~/~
\tau_{\rm SF} \nonumber \\
&\approx& 6\times 10^{11} \left(\delta_{md}\over 5\right)
        \left(10^9\rm yr\over \tau_{\rm SF}\right)
	\left(7.5\over\kappa_{d230}\right) 
        \left(S_{250}\over\rm 4~ mJy\right)L_\odot~
\end{eqnarray}
using Eqs.~(6), (8) and (2). Would this energy be steadily emitted
from the Big Bang to $z\approx 4$, i.e., over a time of $\sim 1$ Gyr,
then the average luminosity would be similar to that of local ULIRGs,
which is bright, but below the detection limit of (sub)millimetre
observations. It appears unlikely that such a high rate of star
formation has been maintained for a Hubble time (however, see Isaak et al. 2001). More likely, vigorous
star formation started at a later time, in a series of short duration
bursts, during which the mass of the heavy elements implied by the
dust emission was produced.

One-third of the PSS quasars studied in this paper
have $L_{\rm FIR} \sim 10^{13} \,\rm  L_\odot$. If most of
this luminosity comes from a starburst, Eq. (9) shows that its
duration cannot much exceed $\rm \sim \, 50 \, Myr$. This is the typical duration
of starbursts in local ULIRGs. The duration of the peak luminosity of
the bright quasars is close to the Eddington-Salpeter time,
$\rm \sim \, 50 \, Myr$ (e.g., Rees 1984). It is likely that most of this
peak-luminosity phase of the PSS quasars that we observed, took place
in the studied redshift range 4--4.5 (which spans a total time
$\rm \sim \,  150 \, Myr$), since few similar objects are known at $z \, > \,4.5$.
Because about one third of the PSS quasars were detected at 1.2~mm, the
duration of their $\rm 10^{13} \, L_\odot$ starburst phase cannot be much
shorter than 15~Myr. It could well exceed this value if similar
starbursts extended over the period when the quasar are less luminous,
as possibly supported by the high detection rate of the low luminosity
part of our sample (Fig.~\ref{figure2}).

In summary, starbursts in high redshift quasars can
consistently account for the far-IR luminosity, the observed dust
content, and the fraction of
quasars which are bright at far-infrared wavelengths.
However, we can not exclude the possibility that the
observed far-IR luminosity is due to AGN heating, and that the dust was
generated in weaker (below our detection limit) and longer (to
generate the required dust mass) past starburst events.

\subsection{Star formation and black hole growth}

Massive black holes are now recognized as a normal, perhaps
ubiquitous, component of elliptical galaxies and spiral galaxy bulges.
There is increasing evidence that the growth of central supermassive
black holes and the formation of the first stellar populations are
closely linked (see, e.g., Combes 2001, and Haehnelt \& Kauffmann 2000 
and references therein). Therefore the presence of strong starbursts in
luminous high-redshift quasars would not be surprising.

In nearby galaxies, there is good evidence for a correlation between  the
black hole mass, $M_{\rm BH}$, and the bulge mass, $M_{\rm bul}$,  with
        \begin{equation}
        M_{\rm bul}/M_{\rm BH} \sim 160-500
        \end{equation}
(Magorrian et al. 1998; Kormendy \& Richstone 1995; Gebhardt et al. 2000
and references therein). Although the average value of this ratio
is close to 500, it appears (see Laor 2001) that
it is lower for the most massive local black holes.

The  $z > 4$ PSS quasars we observed are extremely luminous,
with a typical blue magnitude $M_B = -27.5$ (Table~1).
Adopting the blue luminosity bolometric
correction factor of $\sim 12$ derived for the PG sample by Elvis et
al.  (1994), the typical bolometric luminosity of the quasars is $L_{\rm bol}\rm \approx
10^{14} \, L_\odot$.  Only a fraction, $\epsilon = 0.1\epsilon_{0.1}$,
of the mass equivalent energy accreted onto the black hole is
radiated, so that the rate of growth of a black hole relates to its
luminosity
\begin{equation}
    \dot M_{\rm BH}  ~=~  { L_{\rm bol}\over \epsilon \, c^2 }
                     ~=~  {70~\rm M_\odot yr^{-1}\over
                           \epsilon_{0.1}}~
                            {L_{\rm bol}\rm \over
                           10^{14} \, L_\odot}~.
\end{equation}
The ratio between the star formation rate and the black hole growth
rate is thus
\begin{equation}
  {\dot M_{\rm SF}\over\dot M_{\rm BH}} ~=~
  140 ~\delta_{\rm MF}~ \delta_{\rm SB}~ \epsilon_{0.1}~ (L_{\rm FIR}/L_{\rm
bol})~,
\end{equation}
where we used Eqs.~(4), (5) and (11). The FIR to bolometric luminosity
ratio of the millimetre-{\it detected} PSS quasars ranges between 0.04 and
0.2, with an average value of 0.1, whereas the average ratio over the
{\it whole} sample is $\approx 0.04$. The rate at which stars form in
the PSS quasars is therefore more than a factor ten lower than
expected if stars and black holes formed
simultaneously with the same ratio of bulge
to black hole mass, as observed in
local galaxies.

The fact that the far-IR luminosity in the sample does not appear
to correlate with the blue luminosity (Fig. 2), indicates that the
instantaneous rates of bulge star formation and of black hole growth
are not directly related, and that the rate ratio of Eq. (12) may vary
in time. However, there is a similar relation between the global
masses $M_{\rm bul}/M_{\rm BH}$. The mass of a black hole radiating
near the Eddington limit is (e.g. Rees 1984)
	\begin{equation}
	\rm M_{BH} = 2.6 \times	10^{9} \, \rm M_\odot ~ (L_{\rm bol} \, / \, 10^{14} \, L_\odot).
	\end{equation}
Since the bulge mass can be related to the mass of heavy
elements released (Eq. 7), thus
to the dust mass emitting the observed far-IR radiation
(Eq. 8), which is proportional to the millimetre flux detected (Eq. 2), we find that
        \begin{equation}
        \left(M_{\rm bul}\over M_{\rm BH}\right) \approx  6
        \delta_{\rm MF}\left(\delta_{md}\over 5\right)
	\left(10^{14} L_\odot \over L_{\rm bol}\rm \right)\left(7.5\over\kappa_{d230}\right) 
        \left(S_{250}\over \rm mJy\right)
         \end{equation}
which is independent of the time scale of star formation and of the
source of dust heating. Despite the uncertainty on $\delta_{md}$ and $\kappa_{d230}$, this
ratio is again an order of magnitude smaller on average than the local
observed ratio $M_{\rm bul}$/$M_{\rm BH}$ $\ga$ 160 (Eq. 10).

The bulge to black hole mass ratio which is implied for the present
sample of $z >4$ PSS quasars thus supports the idea that supermassive
black holes grow relatively faster at early epochs than their surrounding host
galaxies. This has also been suggested by observations of host
galaxies of lensed quasars at high $z$ (Rix et al. 1999) and by
modelling (Kauffmann \& Haehnelt 2000).

The relatively faster black-hole growth compared to the bulge growth at $z > 4$
could be in part due to the special conditions of the primordial gas
infall into the central potential wells of very young galaxies and to
the larger proportion of gas in mergings at $z > 4$ than at later
epochs (e.g., Haehnelt \& Rees 1993 and Rees 1998).

Such effects could be enhanced in the PSS quasars which have bolometric
luminosities much higher than those of the average quasar
population. The high-redshift PSS quasars were selected by their
optical brightness, so only the most luminous objects were discovered,
which are not representative of the median $z >$ 4 quasar population
(Fan et al. 2001), nor, a fortiori, of the median $z\approx 0-2$
population (Boyle et al. 2000). The PSS quasars could thus be unusual
objects which at later times have more massive black holes than
typical quasars with comparable bulge mass, or they later accrete a
larger stellar bulge mass without much growth of the black hole. 
 
\subsection{Relation with other quasar properties}

There is as yet very little information published about the PSS quasars
listed in Tables~1 and 2, in particular about their spectra, their
environment and possible lensing. The information from the few
published spectra (Kennefick et al. 1995a,b) is given in Tables~1 and
2. This information is too limited to allow us to confirm that broad
absorption lines (BAL) or/and broad and weak emission lines indicate
strong millimetre emission (Omont et al. 1996a).  Djorgovski (1998,
1999) have observed the surroundings of about twenty such quasars and found
companion galaxies in almost every case. Specific information is only
given for a few sources (as indicated in Tables~1 and 2). These
findings give further support to the general arguments that quasars at
$z > 4$ may form at the very highest peaks of the primordial density
field, in the cores of future giant ellipticals (Djorgovski 1998).
The presence of associated absorption systems points in the same
direction, but we still lack conclusive information on this issue.
Indications that the quasars detected at 1.2~mm have a rich
environment could be related to recent results on high-$z$ radio
galaxies showing spatially distributed dust and CO emission (Ivison et
al. 2000b; Papadapoulos et al. 2000), to the discovery of clusters of
Ly$\alpha$ emission sources (Pentericci et al. 2000) and, if it is not
due to lensing, to the double peak observed in the 1.2~mm dust
continuum of BR B1202$-$0725 (Omont et al. 1996b).

\section{Conclusion}

The availability of ever more sensitive bolometer detectors and
of vast optical surveys have lead to the discovery of thermal
emission from tens of $z > 4$ radio-quiet quasars.
(Sub)millimetre astronomy is thus entering an era of statistical
analysis of star formation in high-$z$ ultra-luminous quasars.
The detection of warm dust with masses $\ga \rm 10^8 \, M_{\odot}$ and
far-infrared luminosities of $\rm \sim 10^{13} \, L_{\odot}$
suggests substantial star
formation activity that could be associated with
a large initial collapse or with major merging.
Millimetre and submillimetre
observations thus provide unique information with which to study the
parallel growths of bulges and central black holes at very large
redshift.  The exceptional objects we discussed may trace the very highest
peaks of the primordial density field, in the cores of proto-clusters
and of future giant ellipticals.

An important question still to be addressed is how the millimetre
emission of bright quasars at $z > 4$ compares to that of quasars in the redshift
range $z \sim 1.5-3$, at which the quasar activity peaks.
In earlier studies (Omont et al. 1996a; Guilloteau et al. 1999)
only a few detections were achieved in this redshift range.

\vspace{1cm}

{\it Acknowledgements:} We thank G. Djorgovski and the DPOSS QSO search
team for making available a list of $z > 4$ PSS quasars on the web. We
also wish to thank R. Priddey for having provided the compilation of
$M_B$ values, and G. Lagache and A. Beelen for useful discussions. We are most grateful to E. Kreysa
and the MPIfR bolometer group for providing MAMBO, to R. Zylka for
creating the MOPSI data reduction package, and to the IRAM staff for
their support.  This work was carried out in the context of EARA, a
European Association for Research in Astronomy. The National Radio
Astronomy Observatory (NRAO) is a facility of the National Science
Foundation, operated under cooperative agreement by Associated
Universities, Inc.

\end{document}